\documentclass[letterpaper,pra,twocolumn,showpacs,superscriptaddress]{revtex4}

\usepackage{graphicx}
\usepackage{amsmath}
\usepackage{amssymb}

\renewcommand{\det}[1]{\textit{det}\left[#1\right]}

\renewcommand{\log}[1]{\text{log}\left[#1\right]}
\newcommand{\tinymath}[1]{\text{\tiny{$#1$}}} 

\DeclareMathOperator{\I}{\textit{I}} 
\DeclareMathOperator{\J}{\textit{J}} 
\DeclareMathOperator{\D}{\mathcal{D}} 

\begin{document}

\title{Experimental Investigation of the Evolution of Gaussian Quantum Discord in an Open System}

\affiliation{Department of Physics, Technical University of Denmark, Fysikvej, 2800 Kgs. Lyngby, Denmark}
\author{Lars S. Madsen}
\noaffiliation{}

\author{Adriano Berni}
\noaffiliation{}

\author{Mikael Lassen}
\noaffiliation{}

\author{Ulrik L. Andersen}
\noaffiliation{}
\email{}

\date{\today}

\begin{abstract}
Gaussian quantum discord is a measure of quantum correlations in Gaussian systems. Using Gaussian discord we quantify the quantum correlations of a bipartite entangled state and a separable two-mode mixture of coherent states. We experimentally analyze the effect of noise addition and dissipation on Gaussian discord and show that the former noise degrades the discord while the latter noise for some states leads to an increase of the discord. In particular, we experimentally demonstrate the near-death of discord by noisy evolution and its revival through dissipation.   
\end{abstract}


\maketitle

Entanglement is undoubtedly a key resource in quantum information science and it has become synonymous with different nonclassical tasks such as quantum teleportation, dense coding and quantum computation~\cite{Nielsen}. However, despite its obvious importance for such tasks, the exact need of entanglement in some other non-classical tasks has remained enigmatic. It has for example been shown that some quantum computational tasks based on a single qubit (the socalled DQC1 model) can be carried out by separable (that is, non-entangled) states that nonetheless carries non-classical correlations~\cite{Laflamme98,Laflamme05,White08}. Another quantum task requiring the use of quantum correlations but not entanglement is quantum key distribution~\cite{BB84,Grangier03}. This indicates that there exist genuine quantum correlations different from entanglement, which allow for information processing that is intractable by classical means.

A measure of non-classicality that also captures the non-classical correlations of separable states is quantum discord as suggested in refs~\cite{Zurek01,Vedral01}. The quest of understanding the possible use of quantum discord for quantum tasks has fueled an explosion in the research of discord spanning from the investigations of its evolution in noisy Markovian~\cite{Giovannetti12a,Giovannetti12b,Paternostro11,Bruss11,Guo10} and non-Markovian~\cite{Feng10,Caldeira10} channels to its role in thermodynamics~\cite{Lutz09} and phase transitions in spin systems~\cite{Sarandy09}. This measure has been given an operational interpretation in terms of the required resource for enabling quantum state merging~\cite{Datta11,Winter11} and local broadcasting~\cite{Piani08}. 

Most work on quantum discord has been focused on finite dimensional systems including qubits and qudits but recently the discussion has been extended to infinite dimensional systems also known as continuous variable systems~\cite{Datta10,Paris10,Korolkova12}. This extension has mainly been restricted to the set of Gaussian states and Gaussian measurements~\cite{Datta10,Paris10}, partly due to the associated reduced mathematical complexity of the problem and partly due to their immense importance for quantum information processing and their simplistic generation~\cite{Andersen10}: Gaussian states can be efficiently prepared~\cite{Kimble86} and allow for secure quantum communication~\cite{Grangier03}, near-optimal cloning~\cite{Andersen05} and teleportation~\cite{Furusawa98}.           

In this Letter we experimentally characterize the Gaussian quantum discord of two-mode squeezed states and separable two-mode mixtures of coherent states. We investigate its evolution in an open quantum system described by local Markovian decoherence in form of Gaussian noise addition corresponding to a classical noise channel and pure attenuation. 

The robustness of non-classical states against noise and loss is of high importance as real-world quantum information protocols will inevitably consist of noisy and lossy operations. Entanglement is known to be a fragile resource which is difficult to generate and the optimal use of it would require complex purification and distillation protocols. On the contrary, it has been shown theoretically that quantum discord of some states is robust against Markovian decoherence, and in fact discord can increase through dissipation both for discrete~\cite{Giovannetti12a,Paternostro11,Bruss11} and for continuous~\cite{Giovannetti12b} variable systems. By interrogating two-mode mixtures of coherent states in an open system we show experimentally that Gaussian quantum discord is reduced in a classically noisy channel and, counterintuitively, increased in a simple dissipative channel. 

\begin{figure}[th]
\begin{center}
\includegraphics[width=0.45\textwidth]{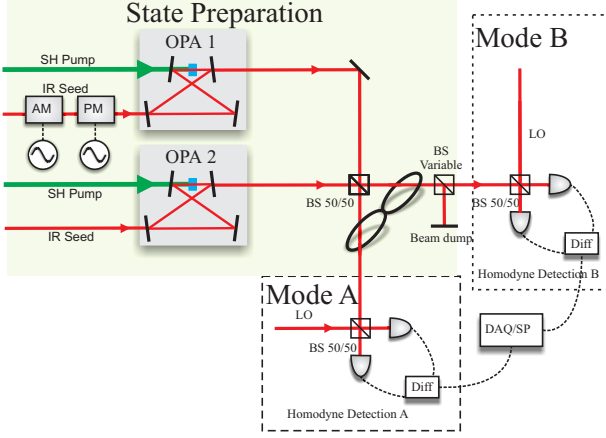}
\caption{Experimental setup. See main text for description. Abbreviations: IR: infrared (1064 nm), SH: second harmonic (532 nm), PM and AM: phase and an amplitude electro optical modulator, OPA: optical parametric amplifier, BS: beamsplitter, LO: local oscillator. DAQ/SP data acquisition/ signal processing}
\label{setup}
\end{center}
\end{figure}

Figure~\ref{setup} shows the experimental setup that was used to generate quantum correlated Gaussian states. A pair of optical parametric amplifiers (OPAs) based on type I quasi-phase-matched periodically poled KTP cystals placed inside bow-tie shaped cavities were used to generate two independent amplitude squeezed beams at 1064 nm. The OPAs were seeded with dim laser beams at 1064 nm to facilitate the locking of the cavities and several phases of the experiment. Both OPA outputs had 3.2 $\pm 0.2$ dB of squeezing and 6.7 $\pm 0.2$ dB of anti-squeezing which was measured with a homodyne detector with a total efficiency of $85 \pm 5 \%$ and electronic noise contribution of $-20$ dB relative to shotnoise. The measurements were performed at the sideband frequency of 4.9 MHz with a bandwidth of 90 kHz. In order to generate correlations in addition to those produced in the OPAs, one of the seed beams was symmetrically modulated with two electro-optic modulators - an amplitude and a phase modulator - that were driven by independent electronic noise generators. We interfered the two squeezed beams on a symmetric beam splitter with a relative and locked phase of $\pi/2$ to produce a pair of quadrature entangled beams. One of the output beams propagated through a dissipative channel which was implemented by a beam splitter with a variable transmittivity. This setup can hence generate and characterize a variety of two-mode Gaussian states. 

Due to the Gaussian nature of the generated two-mode states, they are fully characterised by their covariance matrices which are easily estimated from the homodyne measurements of the amplitude quadrature $X$, and phase quadrature $P$. The covariance matrix for the state $\rho_\tinymath{AB}$ written in the standard form is
\begin{equation}\label{simple_CM}
\sigma_\tinymath{AB} =
 \begin{pmatrix}
  \alpha & \gamma \\
  \gamma^\tinymath{T} & \beta
 \end{pmatrix}
\end{equation}
where $\gamma= \textrm{Diag}\left[\textrm{cov}(x_A, x_B),\textrm{cov}(p_A, p_B)\right]$, $\alpha=\textrm{Diag}\left[ \textrm{var}(x_A),\textrm{var}(p_A)\right]$ and $\beta=\textrm{Diag}\left[\textrm{var}(x_B),\textrm{var}(p_B)\right]$. 

\begin{figure}[th]
\begin{center}
\includegraphics[width=0.45\textwidth]{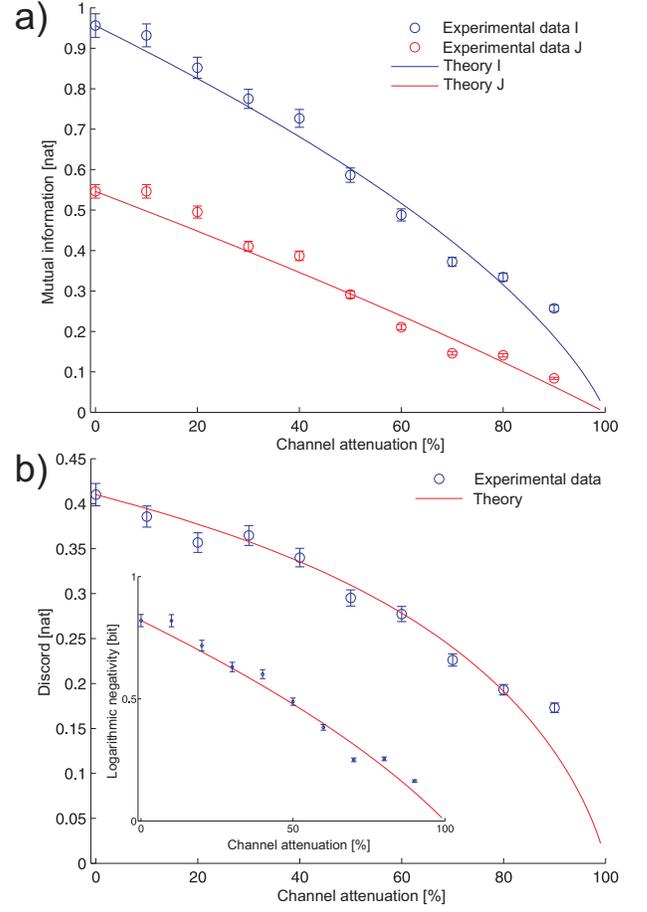}
\caption{a) Mutual information for the two-mode squeezed state with the circles representing the $I$ information (red) and the $J$ information (blue). The curves correpond to theoretical curves fitted to the first data point. 
b) Quantum discord and logaritmic negativity (inset) for the two-mode state as a function of attenuation. The experimental data (circles) show the discord and logaritmic negativity calculated from the measured covariance matrices and the solid lines are theoretical fits to the first measurement point (corresponding to a channel attenuation of 0\%).}
\label{ent}
\end{center}
\end{figure}

Gaussian quantum correlations beyond entanglement are captured by the measure of Gaussian discord which we formalise in the following \cite{Datta10, Paris10}. In a bipartite system, the total amount of correlations (classical and quantum) is given by the von Neumann mutual information $\I(\rho_\tinymath{AB})=S(\rho_\tinymath{A})+S(\rho_\tinymath{B})-S(\rho_\tinymath{AB})$ where $S(\rho)$ is the von Neumann entropy and $\rho_\tinymath{A(B)}$ is the reduced density matrix of the A (B) subsystem. Another measure of mutual information that only quantifies the amount of classical correlations extractable by a Gaussian measurement is  $\J_\tinymath{A}(\rho_\tinymath{AB})=S(\rho_\tinymath{A})-\inf_{\sigma_M}S(\rho_\tinymath{A|\sigma_M})$ where $\sigma_M$ is the covariance matrix of the measurement on mode B. 
As it only captures the classical correlations, the difference, $\D_\tinymath{A}=\I(\rho_\tinymath{AB})-\J_\tinymath{A}(\rho_\tinymath{AB})$, is a measure of Gaussian quantum correlation that is coined Gaussian quantum discord. An explicit expression for this discord has been found~\cite{Datta10}:

\begin{equation}
\D(\sigma_\tinymath{AB})=\mathfrak{f}(\sqrt{I_2})-\mathfrak{f}(\nu_-)-\mathfrak{f}(\nu_+)+\mathfrak{f}(\sqrt{E^\tinymath{min} } )
\label{discord}
\end{equation}
where $\mathfrak{f}(x)=(\frac{x+1}{2})\log{\frac{x+1}{2}}-(\frac{x-1}{2})\log{\frac{x-1}{2}}$ and

\begin{align}
E^\tinymath{min}=\left\{
\begin{array}{l l l}
&\frac{2 I_3^2+(I_2-1)(I_4-I_1)+2|I_3|\sqrt{I_3^2+(I_2-1)(I_4-I_1)}}{(I_2-1)^2} &\textrm{a}) \\
&\frac{I_1 I_2-I_3^2+I_4-\sqrt{I_3^4+(I_4-I_1 I_2)^2-2C^2(I_4+I_1 I_2)}}{2 I_2} & \textrm{b}) \\
\end{array}\right.
\end{align}
where a) applies if $\quad(I_4-I_1 I_2)^2\leq I_3^2(I_2+1)(I_1+I_4)$ and b) applies otherwise. $I_1=\det{\alpha}$, $I_2=\det{\beta}$, $I_3=\det{\gamma}$, $I_4=\det{\sigma_\tinymath{AB}}$ are the symplectic invariants and $\nu_\pm^2=\frac{1}{2}(\delta\pm\sqrt{\delta^2-4I_4})$, with $\delta=I_1+I_2+2I_3$, are the symplectic eigenvalues. 

Using the experimentally obtained covariance matrix of the two-mode squeezed state, we calculate the Gaussian discord for different attenuations of mode B. The results for the mutual informations are shown in Figure~\ref{ent}a and the discord in Figure~\ref{ent}b. As usually expected for quantum correlations in a dissipative channel, we see a monotonic decrease of the Gaussian discord with increasing dissipation. A similar behavior is expected and observed for the entanglement of the state as shown in the inset of Figure~\ref{ent} where the logarithmic negativity is plotted against attenuation~\cite{Bowen03}. 
We thus conclude that for this particular two-mode squeezed state, dissipation has a degrading influence on both entanglement and discord. However, it is well-known that entanglement is not the only resource containing Gaussian discord. A two-mode mixed separable state may also contain Gaussian discord, and as we shall see in the following, this discord is much more resilient to dissipation than the entanglement based discord.

\begin{figure}[th]
\begin{center}
\includegraphics[width=0.45\textwidth]{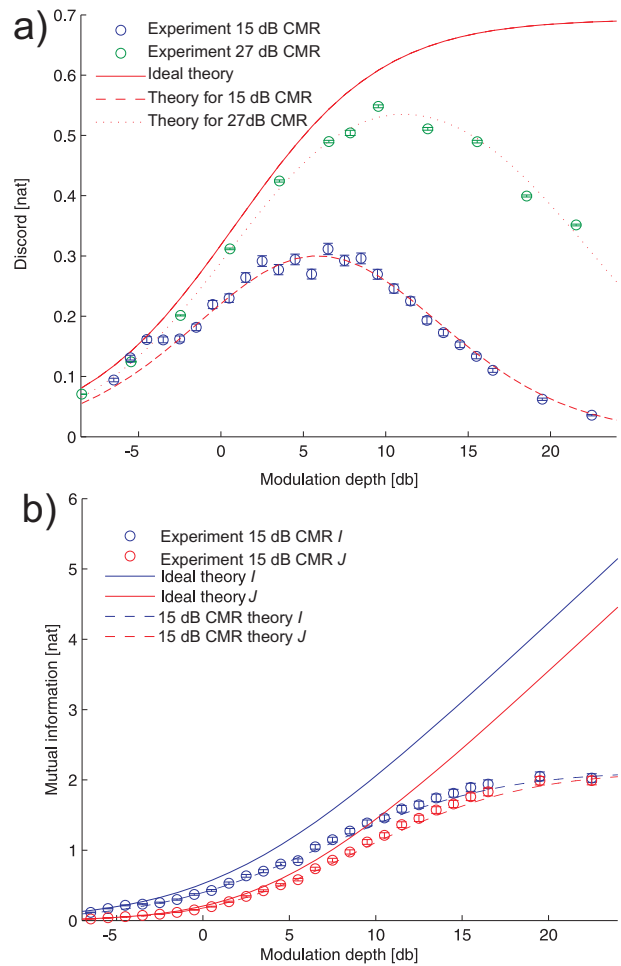}
\caption{a) Quantum discord as a function of the modulation depth. The circles represent the experimentally obtained data for discord with 27~dB (green) and 15~dB (blue) common-mode-rejection (CMR) of the homodyne detectors. b) Mutual informations $I$ and $J$ as a function of the modulation depth. The circles represent the experimentally obtained data. Solid lines result from theory with perfect detectors (corresponding to a noise-free channel) whereas the dashed and dot-dashed curves are the results of the theory including CMR and electronic noise. The error bars are given by the statistical uncertainties.}
\label{modulation}
\end{center}
\end{figure}

In the experiment we generate a separable mixed state by enabling the electronic noise generators and disabling the OPA's (see Figure~\ref{setup}). This corresponds to the splitting of thermal state on a symmetric beam splitter and thus the generation of a two-mode mixture of coherent states with correlations between the amplitude quadratures and the phase quadratures of the two modes. For this state, we measure the covariance matrix for different modulation depths of the modulators and subsequently calculate the discord with the results shown in Figure~\ref{modulation} (green circles). The theoretically expected behavior of the discord for the mixed state is represented by the solid red line which is monotonically increasing, eventually saturating for large modulations. We see that the experimentally obtained values of the discord decrease for very large modulations thus deviating from the theoretically predicted behavior. This discrepancy is simply due to finite balancing between the two homodyne detectors which was measured to have a common mode rejection ratio of 27 dB. Such an imperfection of our detection system leads to the measurement of uncorrelated events which directly simulates a classically noisy channel where the amount  of noise scales with the modulation depth. Since we are interested in simulating the effect of classical noise on the discord, we decided to reduce the common mode rejection ratio of the two detectors to 15 dB. The results for the discord are shown by the red circles in Figure~\ref{modulation}a, and the constituents - the mutual informations $I$ and $J$ - are presented in Figure~\ref{modulation}b. In a classically noisy channel we thus observe the transition from a quantum regime with non-zero discord to a classical regime with near zero discord. 

\begin{figure}[th]
\begin{center}
\includegraphics[width=0.45\textwidth]{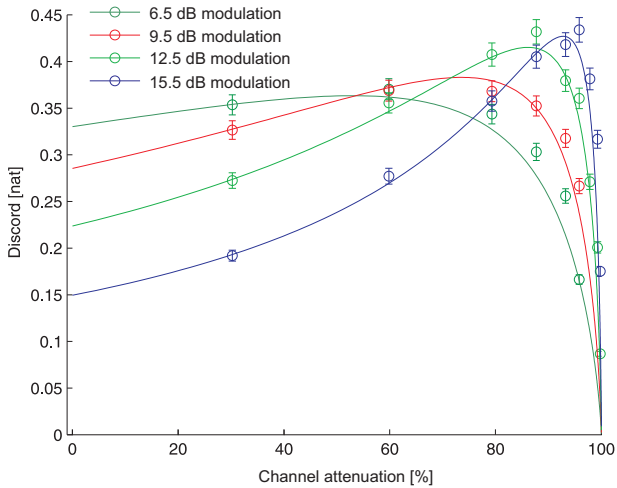}
\caption{The discord of four different mixed states from Figure~\ref{modulation} with different modulation depths as a function of attenuation of mode B. The error-bars are given by the statistical uncertainty. The solid lines are theoretical estimations, fitted to the first data point associated with zero attenuation.}
\label{dissipation}
\end{center}
\end{figure}

In the next step we investigate the effect of dissipation on mode B. Experimentally, dissipation is simulated by a simple beam splitter placed in mode B as shown in Figure~\ref{setup}. The Gaussian discord as a function of channel attenuation is estimated through measurements of the covariance matrix, and the results are shown in Figure~\ref{dissipation} for different amounts of modulation (partially resulting in classically noisy channels). From these measurements we see, interestingly, that the discord increases as a function of local and Markovian dissipation. We do not only observe an extreme robustness of discord associated with the separable state but also a clear rise of quantum correlations during dissipative dynamics. This is in stark contrast to entanglement which cannot increase under any local (unitary or non-unitary) transformations~\cite{Horodecki09}. Ultimately, it is possible to observe the complete death of discord through classical noise addition and its subsequent revival through dissipation as partially demonstrated in Figure~\ref{modulation} (near death) and Figure~\ref{dissipation} (revival). 

We interestingly note that the discord of the classically noisy state investigated above can be revived after death even if the state at A has been measured, stored in a classical memory and subsequently recreated. This is also clear from the fact that a classically noisy channel is theoretically identical to a measurement map~\cite{Piani08}.

The revival of the discord as a result of dissipation can be understood from two effects. The first one is the attenuation of the uncorrelated classical noise in mode B, which is responsible for the near death of the discord in Figure \ref{modulation}. This effect is hence the main contributor to the revival of the discord when uncorrelated noise above shotnoise is present. The second effect that leads to a further increase in the discord results from the relatively higher amplitude of mode A. This makes the inevitable noise contribution of a measurement on mode B more significant and thereby increase the difference between the information obtainable by quantum and classical means. 

\begin{figure}[th]
\begin{center}
\includegraphics[width=0.45\textwidth]{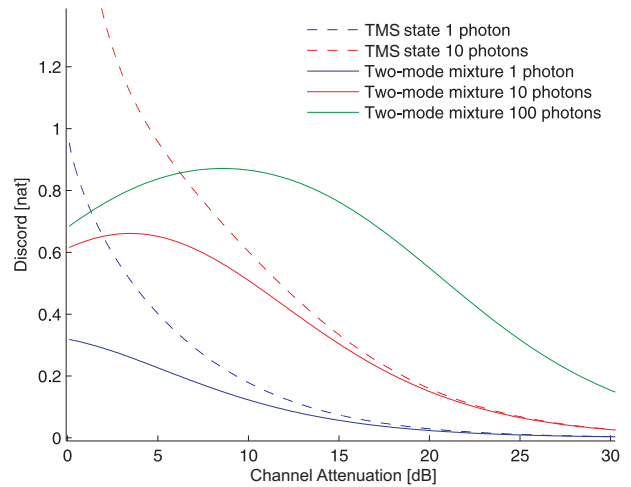}
\caption{Theoretical simulations of the evolution of Gaussian discord in an attenuating channel for pure two-mode squeezed states and two-mode mixtures of coherent states.}
\label{energy}
\end{center}
\end{figure}

Although the Gaussian discord of our experimentally obtained separable states is more resilient to losses than that of the entangled states, it is important to point out that for the same amount of energy of the two types of states, the entangled state will for any amount of Markovian loss exhibit the highest discord. This is illustrated in Figure~\ref{energy} where the discord of pure entangled and separable states with mean photon numbers  between 1 and 100 (1 and 10 photons corresponding to 5.7 dB and 13.4 dB of two-mode squeezing) is plotted against attenuation (in dB). Needless to say, such pure squeezed states are much more difficult to produce than the more robust two-mode mixed coherent state with 100 photons (green curve in Figure~\ref{energy}) which is straightforwardly produced with electro-optical modulators.

Finally, we note that the discord of a combination of entangled states and separable states might be the optimal resource for some applications. E.g. we have recently shown that by exploiting the correlations of combined two-mode squeezed states and coherent mixed states the performance of a continuous variable quantum key distribution system can be improved in terms of increased robustness to channel noise and dissipation~\cite{Madsen11}. 

In conclusion, we have experimentally generated two different states containing Gaussian quantum discord: A two-mode entangled state and a two-mode separable state. The discord associated with the entangled state was shown to degrade under dissipative dynamics whereas the discord of the separable state was exhibiting the opposite behavior; it was an increasing function of dissipation until a certain point after which it rapidly vanished. Furthermore, we show that discord of separable states is reduced if uncorrelated noise is added to the state but that it can be revived through dissipation.
             
We acknowledge the financial support from the Danish Council for Technology and Production Sciences (no. 10-093584 and no. 10-081599).

\end{document}